# Analyzing the Impact of EV Battery Charging on the Distribution Network

Sahil Aziz[1], Wajid Ali[2], Khaliqur Rahman[3]

[1,2,3]Electrical Engineering Department, Aligarh Muslim University, Aligarh, India

*Abstract*— **Many countries are rapidly adopting electric vehicles (EVs) due to their meager running cost and environment-friendly nature. EVs are likely to dominate the internal combustion (IC) engine cars entirely over the next few years. With the rise in popularity of EVs, adverse effects of EV charging loads on the grid system have been observed. Since the distribution system (DS) does not cope with the high overloading capacity, the negative impact of EV charging load on the distribution network (DN) cannot be neglected. A high level of EV penetration with uncoordinated charging is the primary cause of voltage instability, increased peak load demand, and reliability issues of the DN. In this paper, a detailed overview of all the notable impacts of EV charging on voltage profile, power quality, and DS performance is discussed. This work also reviews the different topologies of EV chargers and the issues introduced by power converters on the utility grid. Finally, the strategies for improving the charging of EVs proposed in the literature to consider the random nature of EVs charging, the management of peak loads, and bidirectional power flow are summarized.**

## I. INTRODUCTION

Rising oil prices and environmental issues have prompted people to become more and more interested in clean vehicle technologies such as electric battery vehicles, biodiesel EV, and fuel cell EV. Environmental issues like global warming, the greenhouse effect, and the depletion of fossil fuels are escalating the necessity of electrified transportation to attain a sustainable environment. Despite readily accessible fossil fuels, IC vehicles are being replaced by electric cars expeditiously. Adapting EVs instead of IC engine vehicles has many environmental and economic advantages. However, with the upsurge in EVs, the charging demand also increased, posing several challenges in the power system. Hence, it is necessary to lay the groundwork for charging EVs to replace IC engines completely. There are four charging methods [1–4]:

*(a) Constant Current (CC) charging:* This method provides constant current to charge the vehicle's battery amidst the entire charging process. The high continuous current is applied to achieve fast charging, affecting battery life. Thus, the main challenge is determining an appropriate current limit to minimize the charging time without degrading battery life.

(b) *Constant-Voltage (CV) charging:* In this method, a preset constant voltage is applied to the EV's battery which provides slow charging but extends the battery lifespan. The current gradually drops as the battery starts charging.

(c) *Constant-Current-Constant-Voltage (CC-CV) charging*: The CC-CV charging method is a hybrid method that combines the above two charging methods. There are two charging stages, firstly the battery is charged in CC mode, and after reaching the safe threshold voltage, the charging process switches to CV charging mode. This charging method help to improve both charging times as well as battery Lifespan.

(d) *Multi-stage Constant-Current (MCC) charging*: In this charging method, there are multiple stages of constant current, where the current starts decreasing when the terminal voltage reaches a predetermined threshold. The charging procedure will continue until the battery reaches its maximum capacity. This process needs a complex control algorithm.

Table 1 given below delineates the key parameters and comparison of various charging techniques.

Table.1- Highlights of merits and demerits of different EV charging methods.

| Strategies | Key parameters | Advantages | Disadvantages |
|---|---|---|---|
| Constant current (CC) | *Charging current rate | *It has a limited current capability to avoid overcharging the initial charge. *Easy to install. | *Utilization capacity is low. |
| Constant voltage (CV) | *Charging voltage rate | *It has a limited voltage capability to prevent overvoltage. *Easy to install. | *Causing the lattice collapse of the battery. |
| Constant current-Constant voltage (CC-CV) | *Charging current rate in CC mode. *Charging voltage rate in CV mode. | *Capacity utilization is high. *Terminal voltage is stable. | *Challenge to balance charging speed and temperature variation. |
| Multi-stage constant current (MCC) | *Various CC stages. *Charging current rate is different for each stage. | *Easy to install. *Fast charging | *Challenge to balance charging speed and battery life. |

There are different adverse impacts of EV charging loads on the DS. In [5], the authors present the effects of EV charging loads on a standard DN consisting of 14 buses. It was concluded that the high EV penetration level degraded the transient voltage stability index. In [6], the impact analysis of charging load on the utility grid during off-peak conditions at a fixed location is discussed. A worst-case scenario is the generation of the load close to the peak demand at a different time in the load curve, for instance, when several EVs are simultaneously connected to the distribution line during peak load hours at any critical point. The most optimistic approach for EV-Grid integration is organizing the EV battery charging time so that the EV load is evenly distributed along the load curve and the system's peak load remains unchanged [7]. However, due to the random behavior of EV load, the strategies described above are limitedly successful. In [8], the authors have introduced the modeling of the total charging demand of EVs for probabilistic power flow calculations by considering the critical factors of EV charging behavior at the commercial and residential charging stations. However, this methodology describes the uncontrolled charging of EVs very well but does not include charging improvement strategies. Another probabilistic modeling of EV charging patterns in a residential DN was discussed [9]. In [13], a statistical methodology is adopted, including all factors that may affect the load curve when charging EVs. In [10], a stochastic method for modeling driving patterns of EVs is carried out through monte Carlo simulation. In uncontrolled domestic charging, three significant variables considered were leaving the EV home, reaching the EV home, and the intermediate distance. Later, the charging profile is combined with simulated EV trips to evaluate the contribution to system power.

This paper is organized as follows: *Section – II* delineates the EV charging modes based on speed. *Section – III* demonstrates the suitable converter topology for EV charging and various power converter circuits and their impacts. *Section – IV* presents the effects of EV charging on the voltage profile of the feeder line, daily load curves in a different season, time and areas, and power quality of the DS. *Section–V* presents the harmonic impacts of EV chargers on the various equipment of the power system, and some ways of improving these effects are also discussed. *Section – VI* illustrates the strategies to enhance the effects discussed in the literature.

## II. CHARGING INFRASTRUCTURE OF EV

In general, EV charging is based on two types of chargers i.e., On-board (fabricated inside the vehicle) and Off-board (external) chargers. Depending on the state of charge(SOC) and charging time of the battery, charging of EVs is categorized into four charging modes [5-6]:

*Mode-1:* EVs consist of an onboard charger. The current consumption is low in this mode, and the power injected into the battery is limited. This is the most cost-effective and convenient method to charge EVs at home with no installation cost. Generally, this mode requires 7~15 hours approximately for a fully discharged battery to reach its maximum capacity; thus, it is considered the slowest charging mode.

*Mode-2:* The chargers usually operate at 440V or 230V AC mains not exceeding 32A. This type of charging requires installing high voltage wiring and EVSE units and is usually located in public places. In this mode, the charger takes about 3~8 hours to charge the battery entirely and is more efficient than the mode-1 charger, but the installation cost is high.

*Mode-3:* This mode requires a 3-phase 440V supply and an off-board charger, which charges the vehicle in just a few minutes rather than hours. Thus, it is the fastest charging method for EVs. These charging stations are located in public and commercial locations and are used for electric automobiles and heavy vehicles such as E-buses.

*Mode-4:* Mode 4 charging is known as the DC fast charging as it directly feeds the DC power to the EV battery by converting high voltage ac power into dc. DC fast charging can charge the vehicle's battery from 10% to 80% in 30-60 minutes [13]. The essential advantage of this mode is that the charging time is drastically reduced. It is almost as fast as refueling in a gasoline car; however, it has certain disadvantages, such as it being much more expensive than all the other charging modes. In the commercial area, the peak demand charges are very high.

The installation of charging stations will impose an additional load on the utility grid. In the case of high capacity fast charging stations of EV, the operational parameters of the DN will be reduced. Similarly, the slow charging stations degrades the grid performance as the EV penetration level increases because they draw power from the distribution line, which usually operates at rated load with minimal overloading capacity and low safety limits.

## III. CIRCUIT CONFIGURATION OF EV CHARGER

The converter is one of the most significant components of an EV charger; hence it is essential to analyze the various converters for research. The EV charger developed should have high efficiency, low cost, and the slightest pressure on the system network. High power factor correction is required and minimizes the power losses in the switching elements. The basic EV battery charger comprises of two main stages;
(a) The first stage is AC to DC conversion with power factor correction (PFC).
(b) The second stage is DC to DC conversion, in which the output DC voltage level of the AC-DC converter is converted to the DC voltage level of the battery.

Generally, an EV charger includes a line frequency transformer(LFT) that supplies power to the diode bridge rectifier to form a dc-link voltage fed as the input of the dc-dc buck converter. However, this topology has several disadvantages: voltage drop under different load conditions, overloading, harmonics in grid side current & the line frequency transformers are very expensive & bulky. In the improved design of the EV charger, the LFT is replaced with a high-frequency transformer(HFT), as shown in fig.1 [14]. The HFT provides reactive power control, voltage regulations, galvanic isolation, and easy integration of renewable energy sources (RES), reducing total cost and size considerations. Moreover, it also allows the charger's output stages to be connected in parallel to meet high power requirements [15]. For example, Tesla Supercharger was made by connecting twelve paralleled modules/converters. Now, to make the EV charger compact, a dual active bridge (DAB) converter (with inbuild HFT) is linked at the rectifier's output and feeds the buck converter. Buck converter can improve the output side of the charger by controlling the current & voltage of the circuit.

The diode bridge rectifier is replaced by a PWM rectifier because when a diode bridge rectifier is employed at the input of the EV charger, the charger current will have harmonics, and it might deteriorate the utility grid and add extra heating losses in the equipment of the power system [16]. However, PWM simultaneously manages the reactive power on the grid and provides sinusoidal current with the required power factor(PF) while supplying DC power. Therefore, it can operate as a static VAR compensator (SVC), adjust the power factor of each load, filter out the harmonics in the power line, and significantly improve the power quality of the power DS. This prevents the issues arising due to the factors mentioned above. The topology used for conversion is analogous to the two-stage solid-state transformer topology.

The DAB provides a high power-rating if multiple secondary windings are employed feeding the isolated buck converter

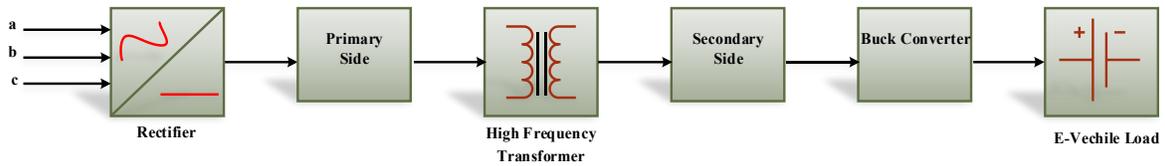

Fig.1: EV Charger with HFT

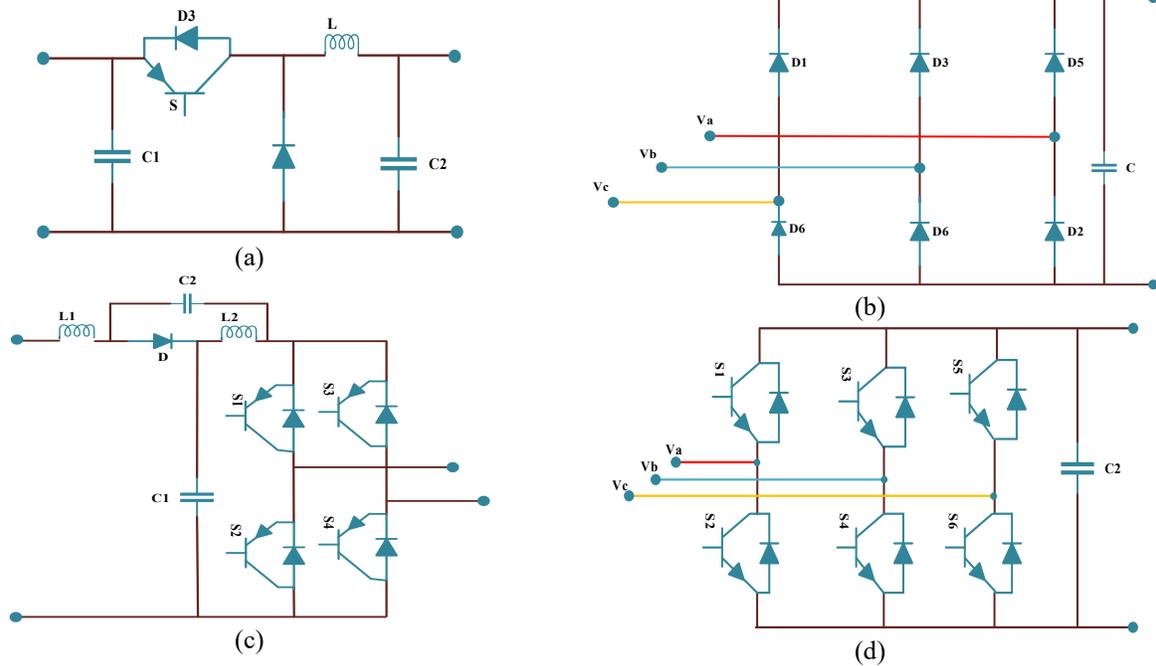

Fig.2- Discrete circuit diagram of (a) Buck Converter (b) Impedance Source Inverter (c) 3-phase Diode Bridge Rectifier (d) 3-Phase PWM Rectifier.

connected in parallel, resulting in a higher charging current at constant voltage.

Two drawbacks of this converter are the high sensitivity of the average active power flow to leakage inductance variation and the high ripple current. Impedance source converter (consisting of Z source and quasi-Z source inverter) due to its smooth start, low switching gate voltage, and low ripple, it is becoming quite popular in EVs [17]. Therefore, to overcome the drawbacks of the DAB converter, impedance source converters are used for both buck and boost operations. The quasi-Z source inverter is an enhanced and upgraded circuit model of the Z-source inverter, with a slightly changed impedance network. This converter topology provides the advantage of additional control achieved by shoot-through zero states. Another advantage is that the circuit has a higher short-circuit strength, which is essential for this converter family. Each converter has its pros and cons, so the impedance source converter has limited boosting potential because of the limited voltage gain during the zero vectors. A Sinusoidal Amplitude High Voltage Bus Converter is employed to suppress the EMI from the circuit to overcome this drawback, resulting in low switching transient period. It can operate adequately with minimum switching loss at high switching frequency while maintaining high power density. Moreover, Multiport dc-dc converter (MPC) is the best option for EVs because it minimizes the input current ripples and output voltage ripples, improves efficiency, and has the potential to handle high power [18]–[20]. Some of the commonly used circuits for power conversion in EV chargers are shown in Fig.2.

## IV. EFFECTS OF EV CHARGING ON DISTRIBUTION SYSTEM

The charging loads of electric vehicle negatively affects the several components as well as the operating parameters of the DS. There are several challenges of non-linear EV charging loads on the distribution network. Fig.3 exhibits the impact on different fundamental parameters and components of DN.

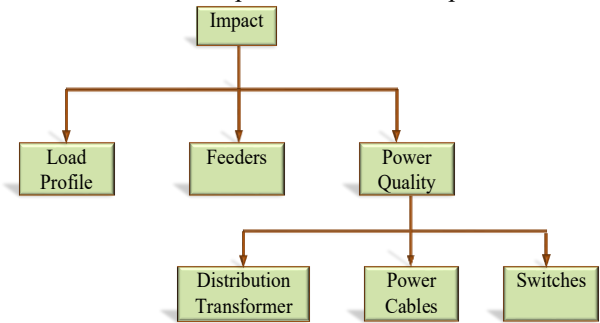

Fig. 3. Impacts of EV charging on Distribution system

### A. Load Profile

The charging behavior of EVs depends on several factors, such as social demographics, time of day, location of charging station, distance, and tax incentives. However, over the year, people keep changing their activities according to the seasons. For example, someone's summer activities may be different from winter activities. Introducing EV's to these two season curves affects in different ways. The ambient temperature in the winter season is low; therefore, the overloading capacity of the conductor increases, whereas the temperature in the summer season is higher, resulting in decreased overloading capacity of the conductor. When the EV charging load is added to these season load curves during peak hours results in increased demand in winters by approximately 16%. Moreover, it is noted that during 11 PM - 6 AM, baseload is low, which creates a significant difference between average load and peak load[21]. During the daytime, photovoltaic modules usually contribute a considerable amount of energy and reduce the burden on the utility grid. However, as dusk arrives, the PV power contribution decreases, and the grid's load increases. During the nighttime, the substantial load increases on the grid conventionally. Furthermore, the EVs are set to charge through the grid[22]. To fulfill this demand, diesel-based generators are integrated with the utility grid. Compared with the baseload plants, the diesel generators have high operating costs and are connected to the grid for a short period, whereas baseload power plants are associated with the grid and put up with most of the network load. Therefore, minimize the peak load of the grid system, and the increased EV load on the grid must be distributed evenly throughout the day [22].

Load density varies with the location. For instance, the urban and suburb load curves are characterized by high customer density, connected to multiple feeders. More priority is given to the suburban/urban areas because the growth of EVs is also more in these areas than in rural areas due to superior financial status. Therefore, the EV load is substantially higher in suburban/urban areas than in rural regions. This difference should be considered when proposing solutions to improve the grid's performance for different parts.

*B. Feeders*

Some challenges are faced by DSs, such as voltage instability, power loss, transformer overloading, etc., which can be avoided by properly planning the DS. Depending on the charging plans and feeder classifications, DS performance changes when EV charging systems are integrated[23]. There are two EV adapting conditions, work dominant & home dominant, that change the load profile of commercial, residential, industrial, and mixed-use feeders. In the case of home-dominant, EV load occurred on the peak time of the non-EV feeder loads, leading to increased load on each feeder, mainly residential load feeders. On the other hand, in the case of work-dominant, charging occurs mostly in the daytime which is generally an off-peak time for non-EV feeder loads. When the non-EV loads are high during the day, the variation in line voltage & line loading is lesser in residential & mixed-use feeders but higher for the commercial feeders [24]. In DS, there are mainly two types of feeders, namely ring main feeder & radial feeder. In radial feeders, voltage dips at the end of the line, whereas in-ring main feeder, voltage dips at the center & stiffs towards the end of the feeder. The introduction of RES and EV loads in the distribution has exacerbated the voltage profile. Under light load conditions, the RES power supply on the PCC can cause overvoltage. In this challenging scenario, as the penetration rate of EV charging increases, the grid voltage curve becomes very unpredictable. Without RES, charging an EV will cause an additional voltage drop on the line and raise the voltage dip for the end user. To overcome this problem, a DT tap switching process is implemented to improve the line voltage distribution. Nevertheless, DT tap changing operation may result in undervoltage or overvoltage at different points of the feeder.

*C. Power Quality:*

Power quality refers to the ability of a power system to provide continuity in service, steady voltages, and currents, smooth sinusoidal waveforms, sustainable power supply within the prescribed limit of voltage, and current harmonics. The common issue with power quality is harmonic distortion, voltage sags, and surges. Electric vehicle charging applications experience power quality issues when the current drawn by the electric vehicle charger consists of harmonics. The deviation of the voltage or current waveform from the ideal sinusoidal waveform leads to harmonic distortion[25]. Current distortion is prevalent for EV charge controllers (non-linear load) due to power electronics switches that convert AC to DC. Feeding these distorted currents into the power DS will distort the utility voltage supply and overload the power distribution equipment. The standard IEEE 519 [26] was formulated to develop "recommended practices and requirements for controlling harmonics in the power grid." Despite frequent observations by manufacturers and researchers, almost all chargers receive highly distorted current from the power system. In the early single-phase diode bridge chargers, the dominant current harmonics were the 3rd and 5th. The three-phase diode charger have the 5th and 7th harmonics as dominant current harmonics. In the case of a thyristor bridge, the dominant current harmonic depends on the number of pulses. For the six-pulse bridge, the significant harmonics are the 3rd, 5th, and 7th.

## V. IMPACT OF HARMONICS ON THE COMPONENTS OF POWER GRID

At the DN level, the utility grid must provide the additional harmonic current drawn by the EV load. Harmonic currents flowing through the DS affect various components such as distribution transformers, power cables, and switches [27].

*Distribution Transformers:* The main effect on the transformer is excessive heating caused by the increased current due to EV battery chargers and losses due to harmonics. An example of the losses due to the increased harmonics in the system is $I^2R$ losses. Other losses due to high harmonic content are stray flux losses.

*Power Cables:* The main effect of harmonics in power cables is the excessive heating caused by increased $I^2R$ losses. This is due to the skin and proximity effect, which depend on the frequency, size, and spacing between the conductors [27], [28].

*Switches:* The presence of harmonic currents may negatively impact switch-gear, protective relaying equipment, metering equipment, and circuit breakers. There is no effect on the behavior of the circuit breaker when interrupting high short-circuit currents. Load distortion will cause higher di / dt at zero crossings, making it more challenging to interrupt in case of light overloads. To eliminate these effects, harmonic filters or harmonic trap filters are used. However, the performance of these filters is limited because their tuning is fixed at the frequency on which the system is designed.

In some cases, the PFC circuit is employed in the EV charger topology to improve the performance of the grid. Earlier, passive power factor correction rectifiers were used in EV stations to reduce harmonics, but they were ineffective as active power filters. Shunt active filters are widely used active filters for EV chargers, which amplifies the harmonics of the load current to reduce the compensation effect of active filters [29]. Though these active & passive power filters have significant advantages for fast and superfast charging stations, the cost of switches & gate drivers for the rated power of EV stations is too high compared to diode bridge rectifiers.

## VI. Optimization Strategies for EV load

Integrating many EVs into the grid network is still a severe issue that should be resolved. The increasing and uncoordinated electrical demand from EV charging poses substantial difficulties to the power grid's ability to operate correctly. However, there are possibilities to optimize the grid performance and make it stable by sticking through some strategies proposed as listed here-

*A. Bidirectional power flow:* Initially, V2G was only concerned with transferring energy from EVs to the DS. However, advancements in technology have added two new power transmission methods (V2H and V2V) [30][31]. An interconnection model of these techniques is shown in Fig.4.

*Vehicle to Grid:* Electrical vehicle batteries can be utilized to provide power to the grid, making the grid more, primarily if vehicle-to-grid (V2G) technology is implemented. V2G technology occurs when a bidirectional EV charger is connected onboard and can transfer EV battery power to the grid. As a result, EVs are both shiftable and distributed energy supplies. V2G allows *EV*s to charge during off-peak hours and return to the grid during peak hours when additional energy is needed. Most of the time, cars remain in the parking area; hence with perfect planning and proper infrastructure, parked EVs could develop powerful energy banks that can stabilize the grid in the future [32], [33].

*Vehicle to Home:* In this, power (electricity) is transferred from the EV battery to the home with the help of a bidirectional EV charger. Likewise, V2G, V2H can also assist the grid during peak hours if the EVs are charged at night when power demand is low and then utilize that energy feeding home at daytime [43].

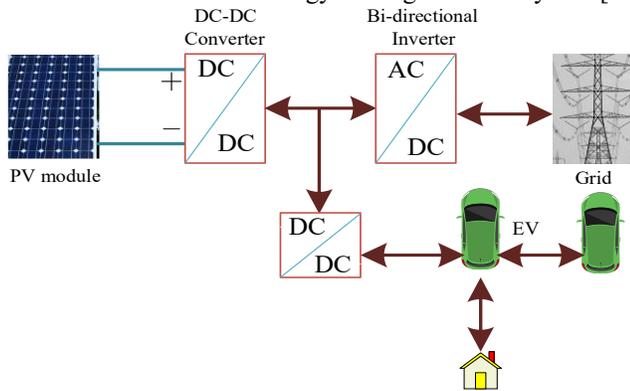

Fig.4: Bidirectional energy transfer

Thus, V2H can ensure that our homes have enough power when they need it the most and minimize the entire grid's burden.

*Vehicle to Vehicle:* In the V2V charging system, power is transferred from one EV to another through a V2V charging module, either wired or wireless. Through this method, cars can be charged anytime, anywhere, just like a powerbank [34]. This charging strategy will immensely reduce the overloads on the DNs, and mobile EV is a feasible solution for the limited range of charging stations.

*B. Coordinated charging:* Coordinated charging, better named smart charging, provides the collective control of charging processes, usually regarded as an essential step towards the effective Grid-EV integration. It allows EVs' charging time to be optimized based on the price or demand of electricity or the generation of renewable energy. The control algorithm and DSOs are used to schedule and decide EV charging, respectively, to uniformly & efficiently distribute the EV load throughout the load curve. Furthermore, it can also minimize the loss in the DN, reduce the DS's investment cost, and minimize the loss of life of the transformer [35].

*C. Dynamic Pricing:* Another strategy for improving the grid performance is dynamic pricing or utility pricing based on usage time. Likewise, smart charging, dynamic pricing is promising to overcome constraints associated with the increased penetration level of EVs. The cost of per unit electricity consumed changes throughout the day; in the morning, when the grid load is minimal or medium, the electricity bill price remains minimal or medium; but under peak load conditions, the consumers pay a high price for this. This will prevent consumers from connecting EV load during high prices and encourage them to use electricity at low prices. This way, reducing the cost of power production, improving grid stability, or improving user satisfaction can be achieved. The customer incentives such as intelligent charging and dynamic pricing are two key factors that unleash the flexibility of EVs, which are essential for the successful integration of EVs into the power grid. A fair representation of dynamic pricing objectives is depicted in Fig.5.

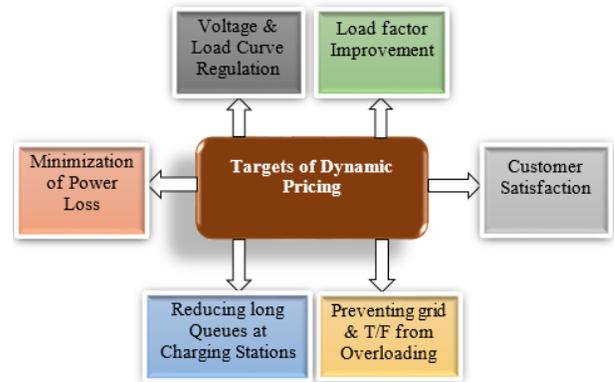

Fig.5: Dynamic pricing

## VII. CONCLUSION

This paper has presented a comprehensive analysis of the impacts of EV charging load on the DS. As the initial contribution, this paper discussed the converter topology commonly used in commercial EV charging stations, which enlisted the problems caused by different EV chargers on the utility grid. Secondly, this work gives depth insight into various impacts on the power quality of the utility grid network and the voltage and load profile of different distribution feeders. Further, problems that occurred due to harmonic currents of EV charger flowing through distribution equipment are discussed. Several effects of EVs' charging behavior on the DS's load curves based on the charging station's area, season, and location are presented. Finally, some strategies are proposed to eliminate these issues and improve the grid's efficiency. Furthermore, this review can provide clear insights to the researchers of optimizing the DS performance and help to establish a roadmap for sustainable expansion of EVs.